\documentclass[preprint,aps]{revtex4}
\usepackage{graphicx}
\usepackage{dcolumn}
\usepackage{bm}
\usepackage{amssymb}
\usepackage{amsmath}

 \newenvironment{Proof.}[1][Proof.]{\begin{trivlist}
     \item[\hskip \labelsep {\bfseries #1}]}{\end{trivlist}}
\begin{document}
\setcounter{page}{1}
\title
{The repulsive nature of naked singularities from the point of
view of Quantum Mechanics}
\author
{D. Batic$^{1,}$$^2$, D. Chin$^{1}$ and M. Nowakowski$^3$}
\affiliation{
$^1$ Department of Mathematics, University of West
Indies, Kingston 6, Jamaica\\
$^2$ Departamento de Matematicas, Universidad de los
Andes, Cra.1E No.18A-10, Bogota, Colombia\\
$^3$ Departamento de Fisica, Universidad de los Andes, Cra.1E
No.18A-10, Bogota, Colombia}

\begin{abstract}
We use the Dirac equation coupled to a background metric to
examine what happens to quantum mechanical observables like the
probability density and the radial current in the vicinity of a
naked singularity of the Reissner-Nordstr\"{o}m type. We find that
the wave function of the Dirac particle is regular in the point of
the singularity. We show that the probability density is exactly
zero at the singularity reflecting quantum-mechanically the
repulsive nature of the naked singularity. Furthermore, the
surface integral of the radial current over a sphere in the
vicinity of the naked singularity turns out to be also zero.
\end{abstract}

\maketitle

\section{Introduction}
The possible existence of naked singularities in General
Relativity (GR) has troubled the physicists and mathematicians
over many years, leading to conjecture that such singularities,
although not forbidden by GR, will be shielded from view by an
event horizon. Such a conjecture is called the cosmic censorship
\cite{CC} and has been proved only on special conditions
\cite{SC}. It was a surprise for the physics community, when
numerical simulations revealed that naked singularity may arise
from gravitational collapse \cite{NS}. This fact brought back on
the agenda of physicists the question how troublesome are naked
singularity really. Sometimes ago it was found that naked
singularities are repulsive in the sense that a particle
experiences a repulsive force in the vicinity of these exotic
objects \cite{RNS}. Here, we examine the issue from the point of
view of Quantum Mechanics. We do so by employing the Dirac
equation in a background metric which contains a naked singularity
of Reissner-Nordstr\"{o}m type. The main result of our
investigation is that neither the wave function nor observables
constructed out of the wave function contain any type of
singularity. It is as if the quantum mechanical particle would not
care about the existence of such an exotic object. This is a
surprising result which is in contrast to the classical case where
the motion of the test particle cannot be continued over the
singularity. Such a result may let think that quantum-mechanically
naked singularities are less troublesome than previously thought.
Furthermore, the numerical value for the probability density at
the singularity turns out to be zero which can be interpreted as
the quantum-mechanical reflection of the classical repulsive
nature of the naked singularity. A result in the very same spirit
concerns the surface integral of the current close to the
singularity which is also zero.

It is worthwhile to recall the unpleasant features of a naked
singularity which seems to allow the existence of closed time-like
curves \cite{timelike}. In the presence of Reissner-Nordstr\"om
and Reissner-Nordstr\"om-deSitter naked singularities this
phenomenon gets reflected in radial velocities $v_r(r)$ of a
massless particle exceeding the speed of light, whenever the
radial distance of the particle from the naked singularity is less
than a certain critical value \cite{wirNEU}. For instance, already
in the Reissner-Nordstr\"om naked singularity scenario we have
$v_r>1$ for $r<Q^2/2M$.

The paper is organized as follows. In section II we will discuss
the Dirac equation in curved space-times. Section III introduces
shortly the angular and radial solutions of the Dirac equation in
the Reissner-Nordstr\"{o}m metric and the main results. Section IV
is devoted to the Klein-Gordon equation in the aforementioned
metric and we show that scalar particles behave similarly to
Fermions in the vicinity of a naked singularity. In section V we
draw our conclusions.

\section{Dirac equation in curved space-times}
To implement the gravity effect into the Dirac equation different
approaches have been used in the literature. One of them is
through the Newman-Penrose formalism \cite{NP}. We follow here the
procedure of \cite{Finster1}, where the Dirac equation is written
in the non-canonical form
\begin{eqnarray*}
&& \left[i\Gamma^\mu(x) D_\mu-m\right]\Psi(x)=0,\\
&& D_\mu=\frac{\partial}{\partial x^\mu}-iE_\mu-ieA_\mu,\\
&&
E_\mu=\frac{i}{2}\rho\partial_\mu\rho-\frac{i}{16}{\rm{Tr}}(\Gamma^\nu\nabla_\mu\Gamma^\lambda)\Gamma_\nu\Gamma_\lambda
+\frac{i}{8}{\rm{Tr}}(\rho\Gamma_\mu\nabla_\lambda\Gamma^\lambda)\rho,\quad
\rho=\frac{i}{4!}\epsilon_{\mu\nu\lambda\sigma}\Gamma^\mu\Gamma^\nu\Gamma^\lambda\Gamma^\sigma
\end{eqnarray*}
where $D_\mu$ is the so-called spin derivative including also
electrodynamics, $\nabla_\mu$ is the covariant derivative w.r.t.
the Christoffel symbols and the gamma matrices satisfy the
anticommutation relations
\[
\{\Gamma^\mu(x),\Gamma^\nu(x)\}=2g^{\mu\nu}(x)
\]
which allow to couple gravity to the Dirac particle. According to
\cite{Finster2}, for every static spherically symmetric space-time
the combination $\Gamma^\mu E_\mu$ simplifies to
\[
B=\Gamma^\mu E_\mu=\frac{i}{2}\nabla_\mu\Gamma^\mu
\]
and the Dirac equation becomes \cite{Finster3}
\[
\left[i\Gamma^\mu\frac{\partial}{\partial
x^\mu}+\frac{i}{2}\nabla_\mu\Gamma^\mu+e\Gamma^\mu
A_\mu\right]\Psi=m\Psi.
\]
The above formulation relies on a local $U(2,2)$ symmetry which
allows a unified description of electrodynamics and general
relativity as a classical gauge theory in the framework of the
Dirac equation \cite{Finster1}. It is therefore possible to choose
a gauge such that
\[
\nabla_\mu\Gamma^\mu=0
\]
bringing the Dirac equation to its canonical form used elsewhere
in the literature \cite{chandra}. We emphasize that such a gauge
fixing is however not necessary in the context of a gauge theory
as any choice of gauge fixing is equivalent. We therefore can also
keep $B$ non zero which leads to a non canonical choice of the
matrices $\Gamma^\mu$. The latter are for the
Reissner-Nordstr\"{o}m line element
\[
ds^2=\Delta(r)dt^2-\frac{dr^2}{\Delta(r)}-r^2(d\vartheta^2+\sin^2{\vartheta}d\varphi^2),
\quad\Delta(r)=1-\frac{2M}{r}+\frac{Q^2}{r^2}
\]
given by \cite{Finster3}
\[
\Gamma^t=\frac{1}{\sqrt{\Delta(r)}}~\gamma^t,\quad
\Gamma^r=\sqrt{\Delta(r)}~\gamma^r,\quad
\Gamma^\vartheta=\gamma^\vartheta,\quad
\Gamma^\varphi=\gamma^\varphi
\]
with
\begin{eqnarray*}
\gamma^t&=&\gamma^0,\\
\gamma^r&=&\gamma^3\cos{\vartheta}+\gamma^1\sin{\vartheta}\cos{\varphi}+\gamma^2\sin{\vartheta}\sin{\varphi},\\
\gamma^\vartheta&=&\frac{1}{r}(-\gamma^3\sin{\vartheta}+\gamma^1\cos{\vartheta}\cos{\varphi}+\gamma^2\cos{\vartheta}\sin{\varphi}),\\
\gamma^\varphi&=&\frac{1}{r\sin{\vartheta}}(-\gamma^1\sin{\varphi}+\gamma^2\cos{\varphi}),
\end{eqnarray*}
where $\gamma^0,\cdots,\gamma^3$ are the Dirac matrices in
Minkowski space-time. The conserved current is
\[
J^\mu=\overline{\Psi}\Gamma^\mu\Psi,\quad\nabla_\mu J^\mu=0
\]
with $\overline{\Psi}=\Psi^\dagger\gamma^0$. This choice of the
adjoint spinor might appear strange is however correct and
justified by the fact that the scalar product
\[
(\Psi|\Phi)=\int_{\mathcal{H}}\overline{\Psi}\Gamma^\mu\Phi~\nu_\mu
d\widetilde{\mu}
\]
with vector field $\nu$ normal to a space-like hypersurface
$\mathcal{H}$ is equivalent to the scalar product derived in the
Newman-Penrose formalism \cite{Batic1} when we integrate over an
hypersurface with $t$ constant and choose the above convention for
the adjoint spinor. To be specific, in the presence of a
Reissner-Nordstr\"{o}m-type naked singularity ($M^2<Q^2$) the
scalar product on the hypersurface with $t$ constant reduces to
\[
(\Psi|\Phi)=\int_{\mathbb{R}^3}\Psi^\dagger\Phi~\frac{d^3
x}{\sqrt{\Delta(r)}}.
\]
We notice that in this formulation $J^0$ can be interpreted as a
positive probability density and $J^i$ with $i=1,2,3$ as the
spatial component of the probability density current. In
particular, $\overline{\Psi}\Gamma^r\Psi$ is the component of the
current in the radial direction.
\section{Solutions of the Dirac equation in a Reissner-Nordstr\"{o}m-type naked background}
By the separation ansatz
\begin{equation}\label{esser}
\Psi(t,r,\vartheta,\varphi)=\frac{1}{r\sqrt{S(r)}}~e^{-i\omega t
}\left(\begin{array}{c}
\Phi^{+}_{jk\omega}(r,\vartheta,\varphi)\\
\Phi^{-}_{jk\omega}(r,\vartheta,\varphi)
\end{array}\right),\quad S(r)=\sqrt{1-\frac{2M}{r}+\frac{Q^2}{r^2}}
\end{equation}
with $M^2<Q^2$ the Pauli spinors $\phi^{\pm}_{jk\omega}$ can be
written as
\[
\Phi^{+}_{jk\omega}(r,\vartheta,\varphi)=\left(\begin{array}{c}
\chi^{k}_{j-\frac{1}{2}}(\vartheta,\varphi)\phi^{+}_{jk\omega,1}(r)\\
i\chi^{k}_{j+\frac{1}{2}}(\vartheta,\varphi)\phi^{+}_{jk\omega,2}(r)
\end{array}\right),\quad
\Phi^{-}_{jk\omega}(r,\vartheta,\varphi)=\left(\begin{array}{c}
\chi^{k}_{j+\frac{1}{2}}(\vartheta,\varphi)\phi^{-}_{jk\omega,1}(r)\\
i\chi^{k}_{j-\frac{1}{2}}(\vartheta,\varphi)\phi^{-}_{jk\omega,2}(r)
\end{array}\right)
\]
where
\begin{eqnarray*}
\chi^{k}_{j-\frac{1}{2}}(\vartheta,\varphi)&=&\sqrt{\frac{j+k}{2j}}~Y^{k-\frac{1}{2}}_{j-\frac{1}{2}}(\vartheta,\varphi)\left(\begin{array}{c}
1\\
0
\end{array}\right)+\sqrt{\frac{j-k}{2j}}~Y^{k+\frac{1}{2}}_{j-\frac{1}{2}}(\vartheta,\varphi)\left(\begin{array}{c}
0\\
1
\end{array}\right),\\
\chi^{k}_{j+\frac{1}{2}}(\vartheta,\varphi)&=&\sqrt{\frac{j+1-k}{2j+2}}~Y^{k-\frac{1}{2}}_{j+\frac{1}{2}}(\vartheta,\varphi)\left(\begin{array}{c}
1\\
0
\end{array}\right)-\sqrt{\frac{j+1+k}{2j+2}}~Y^{k+\frac{1}{2}}_{j+\frac{1}{2}}(\vartheta,\varphi)\left(\begin{array}{c}
0\\
1
\end{array}\right)
\end{eqnarray*}
for $j=\frac{1}{2},\frac{3}{2},\cdots$ and $k=-j,-j+1,\cdots,j$
and $Y^m_\ell$ are the usual spherical harmonics \cite{Thaller}.
The differential equation governing the radial part of the spinor
is \cite{Finster3}
\[
\frac{d}{dr}\Phi^{\pm}_{jk\omega}= \left[ \left(\begin{array}{cc}
0&-1\\
1&0
\end{array}\right)\frac{\omega r-eQ}{rS^2}\pm\left(\begin{array}{cc}
1&0\\
0&-1
\end{array}\right)\frac{2j+1}{2rS}-\left(\begin{array}{cc}
0&1\\
1&0
\end{array}\right)\frac{m}{S}
\right]\Phi^{\pm}_{jk\omega}.
\]
In the vicinity of the naked singularity the radial system takes a
simpler form, namely
\[
\frac{d}{dr}\Phi^{\pm}_{jk\omega}=\pm\alpha_j\left(\begin{array}{cc}
1&0\\
0&-1
\end{array}\right)\Phi^{\pm}_{jk\omega}+\mathcal{O}(r),\quad\alpha_j=\frac{2j+1}{2|Q|}
\]
with the solutions
\begin{equation}\label{asympt_sol}
\phi^{\pm}_{jk\omega,1}(r)\approx e^{\pm\alpha_j r},\quad
\phi^{\pm}_{jk\omega,2}(r)\approx e^{\mp\alpha_j r}.
\end{equation}
The results ensure that the wave functions are regular at the
naked singularity. This enables us to calculate the probability
density $J^0$ in the vicinity of this exotic object. It is a
straightforward computation to show that
\[
J^0(r,\vartheta,\varphi)=\frac{|\Psi|^2}{\sqrt{\Delta}}.
\]
Moreover,
\begin{equation}
J^0(0,\vartheta,\varphi)=0
\end{equation}
since the wave function is
regular at the naked singularity. This result also demonstrates
that the observables constructed out of the wave function are
regular and finite at the singularity. Secondly, the fact that the
probability density is zero at the singularity can be interpreted
in the sense that the singularity has a repulsive nature and this
conclusion is independent of the sign of the combinations of the
two charges $e$ and $Q$. One could say that the particle avoids
the vicinity of the naked singularity from the point of Quantum
Mechanics. Briefly, we also give the results for the radial
component $\overline{\Psi}\Gamma^r\Psi$. The Dirac matrix
$\Gamma^r$ can be conveniently written in the block form
\[
\Gamma^r=\sqrt{\Delta}\left(
\begin{array}{cc}
0&M\\
-M&0
\end{array}
\right),\quad M=\left(
\begin{array}{cc}
\cos{\vartheta}&e^{-i\varphi}\sin{\vartheta}\\
e^{i\varphi}\sin{\vartheta}&-\cos{\vartheta}
\end{array}
\right).
\]
This leads immediately to the expression
\[
J^r_{jk\omega}=\frac{2}{r^2}~{\rm{Re}}(\Phi^{+^{\dagger}}_{jk\omega}M\Phi^{-}_{jk\omega})
\]
where
\[
\Phi^{+^{\dagger}}_{jk\omega}M\Phi^{-}_{jk\omega}=\phi^{+^{*}}_{jk\omega,1}\phi^{-}_{jk\omega,1}
\chi^{k^{\dagger}}_{j-\frac{1}{2}}\chi^{k}_{j+\frac{1}{2}}\cos{\vartheta}
+ie^{-i\varphi}\phi^{+^{*}}_{jk\omega,1}\phi^{-}_{jk\omega,2}|\chi^{k}_{j-\frac{1}{2}}|^2\sin{\vartheta}
\]
\[
-ie^{i\varphi}\phi^{+^{*}}_{jk\omega,2}\phi^{-}_{jk\omega,1}|\chi^{k}_{j+\frac{1}{2}}|^2\sin{\vartheta}-
\phi^{+^{*}}_{jk\omega,2}\phi^{-}_{jk\omega,2}
\chi^{k^{\dagger}}_{j+\frac{1}{2}}\chi^{k}_{j-\frac{1}{2}}\cos{\vartheta}.
\]
Let us now consider a surface integral
\[
\mathcal{S}_{{ }_{B(R)}}=\int_{B(R)}dA~\widehat{{\bf n}}\cdot{\bf
J}=\frac{2}{R^4}~{\rm{Re}}\int
d\Omega~\Phi^{+^{\dagger}}_{jk\omega}M\Phi^{-}_{jk\omega}
\]
where $S$ is a sphere of fixed radius $R$ around the naked
singularity. In general, we have
\[
\int
d\Omega~\Phi^{+^{\dagger}}_{jk\omega}M\Phi^{-}_{jk\omega}=\frac{k}{2j(j+1)}
\left(\phi^{+^{*}}_{jk\omega,1}\phi^{-}_{jk\omega,1}-\phi^{+^{*}}_{jk\omega,2}\phi^{-}_{jk\omega,2}\right).
\]
Specializing now to a tiny region around the singularity with a radius $\epsilon$ we can
use the asymptotic wave functions given in (\ref{asympt_sol}) to
show that
\begin{equation}
\mathcal{S}_{{ }_{B(\epsilon)}}=0
\end{equation}
 This result does not hold
necessarily if the sphere on which we integrate is far from the
singularity. We can interpret $\mathcal{S}=0$ as follows: the
number of ingoing and outgoing particles per unit time through the
surface of the sphere is zero. We think that this again indicates
the quantum-mechanical nature of repulsivity of a naked
singularity. More intuitively, the particle does not dwell for a
long period in the vicinity of the singularity and in particular,
it will not be absorbed.

\section{The Klein-Gordon case}
It is illustrative to have a glimpse on the wave function of
Klein-Gordon particles in the background of a
Reissner-Nordstr\"{o}m naked singularity. In such a geometry the
Klein-Gordon equation
\[
(\square^2-m^2)\Phi=0,\quad\square^2=\frac{1}{\sqrt{-g}}~\partial_\mu(\sqrt{-g}~g^{\mu\nu}\partial_\nu)
\]
leads to the equation
\[
\frac{1}{S^2}~\partial_{tt}\Phi-\frac{1}{r^2}~\partial_{r}(r^2S^2~\partial_{r}\Phi)
-\frac{1}{r^2\sin{\vartheta}}~\partial_\vartheta(\sin{\vartheta}~\partial_\vartheta\Phi)-
\frac{1}{r^2\sin^2{\vartheta}}\partial_{\varphi\varphi}\Phi-m^2\Phi=0
\]
which can be separated by means of the ansatz
\[
\Phi(t,r,\vartheta,\varphi)=e^{-i\omega t}R_{\ell
k\omega}(r)Y^{k}_{\ell}(\vartheta,\varphi)
\]
where $\ell=0,1,2,\cdots$, $k=-\ell,-\ell+1,\cdots,\ell$ and $S$
given by (\ref{esser}). In the naked singularity regime the
equation governing the radial part of the wave function reads
\[
\frac{d}{dr}\left(r^2 S^2\frac{d}{dr}R_{\ell
k\omega}\right)+\left[\frac{\omega^2 r^2}{S^2}+m^2
r^2-\ell(\ell+1)\right]R_{\ell k\omega}=0.
\]
In the vicinity of the naked singularity the radial equation
simplifies as follows
\[
\left[\frac{d^2}{dr^2}-\frac{\ell(\ell+1)}{Q^2}\right]R_{\ell
k\omega}\approx 0
\]
with the solutions
\begin{equation}\label{asympt_sol2}
R_{jk\omega,1}(r)\approx e^{\alpha_\ell r},\quad
R_{jk\omega,2}(r)\approx e^{-\alpha_\ell
r},\quad\alpha_\ell=\frac{\sqrt{\ell(\ell+1)}}{|Q|}.
\end{equation}
The above results ensure that the wave functions are regular at
the naked singularity. We introduce the inner product \cite{Car}
\begin{equation}\label{2stellaQFTCS}
(\Phi|\Psi)=i\int_{\Sigma_t}d^3x~n^\mu\sqrt{-\gamma}~(\Phi\nabla_\mu\Psi^{*}-\Psi^{*}\nabla_\mu\Phi)
\end{equation}
where $\Sigma_t$ is a space-like hypersurface, $\gamma_{ij}$ is
the induced metric on $\Sigma_t$ and $n^\mu$ is a time-like unit
vector normal to $\Sigma_t$. Moreover,
\[
J^\mu_{KG}=-i(\Phi\nabla^\mu\Phi^{*}-\Phi^{*}\nabla^\mu\Phi)
\]
is a conserved current and its zero component is positive definite
only if restricted to positive frequency solutions of the
Klein-Gordon equation (according to the time-like vector field
$t$). However, in the case of a naked singularity of the
Reissner-Nordstr\"{o}m type the vector field $t$ is time-like for
all $r>0$. It is a straightforward computation to show that
\[
J^0_{KG}(r,\vartheta,\varphi)=2\omega\frac{|\Phi|^2}{\Delta}.
\]
As the wave
function is regular at the naked singularity, we obtain
\begin{equation}
J^0_{KG}(0,\vartheta,\varphi)=0
\end{equation}
Finally, we proceed
as we did for the Dirac equation and the radial component of the
current is given by the expression
\[
J^r_{\ell
k\omega,KG}=\frac{2}{r^2}~\delta(r)~|Y^{k}_\ell|^2~{\rm{Im}}(R_{\ell
k\omega}\partial_r R^{*}_{\ell k\omega}),\quad
\delta(r)=r^2-2Mr+Q^2.
\]
We consider again a surface integral
\[
\mathcal{S}_{{ }_{B(R)}}=\int_{B(R)}dA~\widehat{{\bf n}}\cdot{\bf
J}=\frac{2}{R^4}~\delta(R)~{\rm{Im}}\int d\Omega~|Y^{k}_\ell|^2
R_{\ell k\omega}\partial_r R^{*}_{\ell
k\omega}=\frac{8\pi}{R^4}~\delta(R)~{\rm{Im}}~(R_{\ell
k\omega}\partial_r R^{*}_{\ell k\omega})
\]
where $S$ is a sphere of fixed radius $R$ around the naked
singularity. With $\epsilon$ a tiny radius around the
singularity we can use the asymptotic wave functions given in
(\ref{asympt_sol2}) to proof that
\begin{equation}
\mathcal{S}_{{ }_{B(\epsilon)}}=0
\end{equation}
since in
that region the quantity $R_{\ell k\omega}\partial_r R^{*}_{\ell
k\omega}$ is purely real
\[
R_{\ell k\omega}\partial_r R^{*}_{\ell
k\omega}=|c_1|^2e^{2\alpha_\ell r}-|c_2|^2e^{-2\alpha_\ell
r}+2{\rm{Re}}(c_1c_2^{*}).
\]
\section{Conclusions}
There exists a class of particular quantum effects with background
gravity coupled quantum-mechanically to a relativistic particle,
say via the Dirac or Klein-Gordon equation. The peculiarity of
such effects  is their sharp contrast to the results obtained in
the classical framework. One such effect is the absence of bound
states in the Dirac equation coupled to an exterior black hole
metric \cite{boundstates, Finster3} which is in contrast to the
classical result of a stable bound orbit. Another such effect of
background gravity is discussed in the present work. Classically,
the mere existence of a naked singularity poses many problems,
among other the fact that classically the motion of the test
particle cannot be continued over the singularity. We showed that
the situation is less problematic in the quantum mechanical
framework. Indeed, the solutions of the Dirac and Klein-Gordon
equation coupled to the Reissner-Nordstr\"om metric with a naked
singularity reveals that the wave-function is regular at the
singularity. Furthermore, the probability density at the
singularity is exactly zero and the surface integral over the
probability current close to the singularity is also vanishing.
This demonstrates that observables constructed from the weave
 function do not get affected by the presence of the singularity
and their value can be continued up to the singularity itself. The
above findings can be also interpreted as the quantum mechanical
manifestation of the repulsive nature of a naked singularity. From
the point of view of quantum mechanics the particle prefers not to
dwell close to the singularity. In addition results concerning the
quantum mechanical current reveal that the number of particles
entering a sphere close to the singularity per unit time is equal
to the number of particles leaving the sphere per unit time.

In future, it might be a worthwhile undertaking to construct a
velocity operator along the lines explained in \cite{greiner} in
order to determine if the expectation value of the velocity
operator takes values bigger than one.

\end{document}